\documentclass[aps,prl,reprint]{revtex4-1}%
\usepackage{amssymb}
\usepackage{amsfonts}
\usepackage{amsmath}
\usepackage{graphicx}%
\providecommand{\U}[1]{\protect\rule{.1in}{.1in}}

\begin{document}
\preprint{INL/MIS-10-19843 }
\title[ ]{Iterative method for solution of radiation emission/transmission matrix equations}
\author{Clinton DeW. Van Siclen}
\email{clinton.vansiclen@inl.gov, cvansiclen@gmail.com}
\affiliation{Idaho National Laboratory, Idaho Falls, Idaho 83415, USA}
\keywords{}
\pacs{}

\begin{abstract}
An iterative method is derived for image reconstruction. Among other
attributes, this method allows constraints unrelated to the radiation
measurements to be incorporated into the reconstructed image. A comparison is
made with the widely used Maximum-Likelihood Expectation-Maximization (MLEM) algorithm.

\end{abstract}
\volumeyear{year}
\volumenumber{number}
\issuenumber{number}
\eid{identifier}
\date{4 January 2011}
\startpage{1}
\endpage{102}
\maketitle

Imaging by radiation emission or transmission effectively produces a set of
linear equations to be solved. For example, in the case of coded aperture
imaging, the solution is a \textquotedblleft reconstructed\textquotedblright%
\ set of radiation sources, while in the case of x-ray interrogation, the
solution is a set of attenuation coefficients for the voxels comprising the
volume through which the x-ray beam passes.

The linear equations have the form%
\begin{equation}
d_{i}=\sum\limits_{j=1}^{J}M_{ij}\mu_{j} \label{e1}%
\end{equation}
where the set $\left\{  d_{i}\right\}  $ corresponds to the radiation
intensity distribution recorded at a detector (a detector pixel is labeled by
the index $i$), the set $\left\{  \mu_{j}\right\}  $ is the solution, and the
matrix element $M_{ij}$ connects the \textit{known} $d_{i}$ to the
\textit{unknown} $\mu_{j}$. Typically the matrix $M$\ is non-square so that
$\left\{  \mu_{j}\right\}  $ cannot be obtained by standard matrix methods.
(And note that, when the set of equations is large, it can be difficult to
ascertain \textit{a priori} whether the equation set is over- or under-determined.)

In any case the matrix equation $d=M\mu$ may be solved by the iterative method
that is derived as follows. Clearly this method will feature a relation
between $\mu_{j}^{(n)}$ and $\mu_{j}^{(n-1)}$, where $n$ is the iteration
number. Consider the two equations for $\mu_{j}^{(n)}$ and $\mu_{j}^{(n-1)}$,%
\begin{equation}
d_{i}^{(n)}=%
{\textstyle\sum\nolimits_{j}}
M_{ij}\mu_{j}^{(n)} \label{e2}%
\end{equation}%
\begin{equation}
d_{i}^{(n-1)}=%
{\textstyle\sum\nolimits_{j}}
M_{ij}\mu_{j}^{(n-1)} \label{e3}%
\end{equation}
and rewrite the latter as%
\begin{equation}
d_{i}=\frac{d_{i}}{d_{i}^{(n-1)}}%
{\textstyle\sum\nolimits_{j}}
M_{ij}\mu_{j}^{(n-1)}\text{.} \label{e4}%
\end{equation}
Then the relationship between $\mu_{j}^{(n)}$ and $\mu_{j}^{(n-1)}$ is
obtained by setting $%
{\textstyle\sum\nolimits_{i}}
d_{i}^{(n)}=%
{\textstyle\sum\nolimits_{i}}
d_{i}$:%
\[%
{\textstyle\sum\nolimits_{i}}
\left(
{\textstyle\sum\nolimits_{j}}
M_{ij}\mu_{j}^{(n)}\right)  =%
{\textstyle\sum\nolimits_{i}}
\left(  \frac{d_{i}}{d_{i}^{(n-1)}}%
{\textstyle\sum\nolimits_{j}}
M_{ij}\mu_{j}^{(n-1)}\right)
\]%
\[%
{\textstyle\sum\nolimits_{j}}
\left\{  \mu_{j}^{(n)}%
{\textstyle\sum\nolimits_{i}}
M_{ij}\right\}  =%
{\textstyle\sum\nolimits_{j}}
\left\{  \mu_{j}^{(n-1)}%
{\textstyle\sum\nolimits_{i}}
\left(  \frac{d_{i}}{d_{i}^{(n-1)}}M_{ij}\right)  \right\}
\]%
\begin{equation}
\mu_{j}^{(n)}=\mu_{j}^{(n-1)}\frac{1}{%
{\textstyle\sum\nolimits_{i}}
M_{ij}}%
{\textstyle\sum\nolimits_{i}}
\left(  \frac{d_{i}}{d_{i}^{(n-1)}}M_{ij}\right)  \text{.} \label{e5}%
\end{equation}
Note that this last equation can be written%
\begin{equation}
\mu_{j}^{(n)}=\mu_{j}^{(n-1)}\frac{\left\langle \frac{d_{i}}{d_{i}^{(n-1)}%
}M_{ij}\right\rangle _{i}}{\left\langle M_{ij}\right\rangle _{i}} \label{e6}%
\end{equation}
where the last factor is essentially a weighted average of all $d_{i}%
/d_{i}^{(n-1)}$. Thus the set $\left\{  \mu_{j}^{(n)}\right\}  $ approaches a
solution $\left\{  \mu_{j}\right\}  $ by requiring $%
{\textstyle\sum\nolimits_{i}}
d_{i}^{(n)}=%
{\textstyle\sum\nolimits_{i}}
d_{i}$ at each iteration; in effect, by requiring all $d_{i}^{(n)}\rightarrow
d_{i}$.

The iteration procedure alternates between use of Eq. (\ref{e3}) and Eq.
(\ref{e5}) until all $d_{i}^{(n)}$ are as close to $d_{i}$ as desired. For the
first ($n=1$) iteration, an initial set $\left\{  \mu_{j}^{(0)}\right\}  $ is
chosen, which produces the set $\left\{  d_{i}^{(0)}\right\}  $ according to
Eq. (\ref{e3}). These values are used in Eq. (\ref{e5}), so producing the set
$\left\{  \mu_{j}^{(1)}\right\}  $. And so on\ldots\ That a final set
$\left\{  \mu_{j}^{(n)}\right\}  $ is a solution $\left\{  \mu_{j}\right\}  $
to the matrix equation $d=M\mu$\ is verified by checking that all $d_{i}%
^{(n)}=d_{i}$ to within a desired tolerance.

Some cautions and opportunities follow from this simple derivation of Eq.
(\ref{e5}). A caution is that, in the event the equation set is \textit{under}%
-determined, different initial sets $\left\{  \mu_{j}^{(0)}\right\}  $ will
lead to different final sets $\left\{  \mu_{j}\right\}  $ that satisfy the
matrix equation. The corresponding opportunity is that this problem may be
mitigated to some extent by the addition, to the original set of equations, of
linear equations that further constrain the $\mu_{j}$ (perhaps derived from,
for example, independent knowledge of some of the contents of a container
under interrogation). In general the $d_{i}$ appearing in a constraint
equation will have nothing to do with radiation intensity.

The form of any added constraints, and the initial choice $\left\{  \mu
_{j}^{(0)}\right\}  $, must allow all $\mu_{j}^{(n)}\rightarrow\mu_{j}$ and
$d_{i}^{(n)}\rightarrow d_{i}$ monotonically. In particular, care should be
taken when a constraint has one or more coefficients $M_{ij}<0$, as that
affects the denominator $%
{\textstyle\sum\nolimits_{i}}
M_{ij}$ in Eq. (\ref{e5}) (a straightforward fix may be to reduce the
magnitudes of all $M_{ij}$ coefficients and $d_{i}$ in that constraint
equation by a multiplicative factor). In any event, the acceptability of a
constraint equation is easily ascertained by monitoring the behavior
$d_{i}^{(n)}\rightarrow d_{i}$ for that constraint.

Note that \textit{all} solutions $\left\{  \mu_{j}\right\}  $ to a set of
equations that includes additional constraints with $d_{i}>0$ and all
$M_{ij}\geq0$ are accessible from sets $\left\{  \mu_{j}^{(0)}\right\}  $ of
initial values, and further that \textit{any} set $\left\{  \mu_{j}%
^{(0)}\right\}  $ will produce a solution $\left\{  \mu_{j}\right\}  $. This
suggests that, for this implementation of constraints, a superposition of many
solutions may give a good \textquotedblleft probabilistic\textquotedblright%
\ reconstruction. To achieve this, consider that the innumerable solutions to
the set of equations may be regarded as points in a $J$-dimensional space ($J$
is the number of elements in a solution $\left\{  \mu_{j}\right\}  $). These
points must more-or-less cluster, producing a cluster \textit{centroid} that
is itself a solution. While the centroid solution $\left\{  \mu_{j}%
^{(c)}\right\}  $ has no intrinsic special status (as all cluster points
represent equally likely reconstructions), it may be taken to
\textit{represent} the particular set of equations. The cluster size, which
indicates the degree to which solutions are similar to one other, should
decrease as constraints are added. A logical measure of the cluster size is%
\begin{equation}
\sigma_{\text{cluster}}=\left\langle \left(  \mathbf{x}_{c}-\mathbf{x}%
_{k}\right)  \cdot\left(  \mathbf{x}_{c}-\mathbf{x}_{k}\right)  \right\rangle
_{k}^{1/2} \label{e7}%
\end{equation}
where $\mathbf{x}_{c}$ is the centroid vector and $\mathbf{x}_{k}$ is the
vector corresponding to the \textit{k}th solution. Thus the quantity
$\sigma_{\text{cluster}}/\sqrt{J}$, which represents the standard deviation of
the innumerable values of an arbitrary element $\mu_{j}$, is a useful measure
of the variation among solutions $\left\{  \mu_{j}\right\}  $. In general it
is desirable that the variation \textit{among} solutions be much less than the
variation \textit{within} the centroid solution, which is%
\begin{equation}
\sigma_{\mu}^{(c)}=\left\langle \left(  \mu_{j}^{(c)}-\overline{\mu^{(c)}%
}\right)  ^{2}\right\rangle _{j}^{1/2} \label{e8}%
\end{equation}
where $\overline{\mu^{(c)}}=\left\langle \mu_{j}^{(c)}\right\rangle _{j}$. In
that case ($\sigma_{\text{cluster}}J^{-1/2}\ll\sigma_{\mu}^{(c)}$) the
centroid solution is little changed by additional constraints, so suggesting
that the centroid solution $\left\{  \mu_{j}^{(c)}\right\}  $ may be regarded
as the sought-after reconstruction.

Another caution follows from the fact that the denominator $%
{\textstyle\sum\nolimits_{i}}
M_{ij}$ in Eq. (\ref{e5}) is the sum of all elements in column $j$ of matrix
$M$. This iterative method can of course be used without explicitly converting
a set of linear equations into a matrix equation (or \textit{several} sets of
equations into a \textit{single} matrix equation), but in that event very
careful attention must be paid to get the factors $%
{\textstyle\sum\nolimits_{i}}
M_{ij}$ right.

It may be noticed that Eq. (\ref{e5}) is similar to the so-called
Maximum-Likelihood Expectation-Maximization (MLEM) algorithm (see refs.
\cite{r1} and \cite{r2} for derivations of the latter, and see numerous papers
in the recent imaging literature for applications of it). The MLEM, which is
derived from physical considerations having to do with radiation emission and
detection, purports to find the set $\left\{  \mu_{j}\right\}  $ that
\textit{maximizes} the probability $P\left(  \left\{  d_{i}\right\}  |\left\{
\mu_{j}\right\}  \right)  $, which is the probability of realizing the
observed set $\left\{  d_{i}\right\}  $ given a set $\left\{  \mu_{j}\right\}
$. This is in contrast to the derivation above, which leads to Eq. (\ref{e5})
as simply a method to find a solution to a matrix equation (a set of linear
equations). The derivation presented here makes clear how the iterative
procedure should be implemented for an application, and allows constraints to
be added to the original set of equations (those produced by the imaging
exercise) thereby enabling a more-accurate reconstruction when the original
equation set is under-determined.

This work was supported in part by the INL Laboratory Directed Research and
Development Program under DOE Idaho Operations Office Contract DE-AC07-05ID14517.

\end{document}